\documentclass{pasj00}

\begin{document}
\SetRunningHead{Suwa et al.}{Explosion geometry of supernovae}

\title{Explosion geometry of a rotating 13 $M_{\odot}$ star driven by
  the SASI-aided neutrino-heating supernova mechanism}

\author{
Yudai \textsc{Suwa}\altaffilmark{1},
Kei \textsc{Kotake}\altaffilmark{2,3}, 
Tomoya \textsc{Takiwaki}\altaffilmark{3}, 
Stuart C. \textsc{Whitehouse}\altaffilmark{4},
Matthias \textsc{Liebend\"orfer}\altaffilmark{4},
and
Katsuhiko \textsc{Sato}\altaffilmark{5,6}
}
\altaffiltext{1}
{Department of Physics, School of Science,
The University of Tokyo, Tokyo 113-0033}
\altaffiltext{2}
{Division of Theoretical Astronomy, National
  Astronomical Observatory of Japan, Mitaka, Tokyo 181-8588, Japan}
\altaffiltext{3}
{Center for Computational Astrophysics, National
  Astronomical Observatory of Japan, Mitaka, Tokyo 181-8588, Japan}
\altaffiltext{4}
{Department of Physics, University of Basel,
  Klingelbergstr. 82, CH-4056 Basel, Switzerland}
\altaffiltext{5}
{The Institute for the Physics and Mathematics of the
  Universe, the University of Tokyo, Kashiwa, Chiba, 277-8568, Japan}
\altaffiltext{6}
{Department of Physics, School of Science and
  Engineering, Meisei University, 2-1-1 Hodokubo, Hino-shi, Tokyo
  191-8506, Japan}

\email{suwa@utap.phys.s.u-tokyo.ac.jp, suwa@yukawa.kyoto-u.ac.jp}


\maketitle

\begin{abstract}
By performing axisymmetric hydrodynamic simulations of core-collapse
supernovae with spectral neutrino transport based on the isotropic
diffusion source approximation scheme, we support the assumption that
the neutrino-heating mechanism aided by the standing accretion shock
instability and convection can initiate an explosion of a 13
$M_{\odot}$ star.  Our results show that bipolar explosions are more
likely to be associated with models which include rotation.  We point
out that models, which form a north-south symmetric bipolar explosion,
can lead to larger explosion energies than for the corresponding
unipolar explosions.
\end{abstract}

\section{Introduction}

Core-collapse supernovae have long attracted the attention of
astrophysicists because they have many facets playing important roles
in astrophysics. They herald the birth of neutron stars and black
holes; they are a major site for nucleosynthesis; they influence
galactic dynamics; they trigger further star formation and they are
prodigious emitters of neutrinos and gravitational waves. Despite
rigorous theoretical studies for more than 40 years, the details of
the explosion mechanism have been obscured under the thick veils of
massive stars.

For more than two decades, the neutrino-heating mechanism
\citep{wils85,beth85}, relying on the energy deposition via neutrinos
behind the stalled shock, has been supposed as the most promising
scenario.  However, one important lesson we have learned from the work
of \citet{lieb01,ramp02,thom03,sumi05} is that the neutrino heating,
albeit with the best input physics and numerics to date, fails in
spherical symmetry (1D) (see, however, \cite{kita06}).

Pushed by supernova observations of the blast morphology (e.g.,
\cite{wang01,tana07}), it is now almost certain that the breaking of
the spherical symmetry is the key to the supernova puzzle.  The
multi-dimensional (multi-D) hydrodynamic motion associated with
convective overturn in the postshock region
\citep{hera94,burr95,jank96,frye02,frye04b} and the recently
identified standing accretion shock instability (SASI) (e.g.,
\cite{blon03,sche04,ohni06,fogl07,murp08,iwak08,iwak09,guil09}), are
expected to help the neutrino-driven explosion mechanism. This is
because the sojourn time of the accreting material in the gain region
can be longer than in the 1D case, which enhances the efficiency of
the energy deposition behind the stalled shock.

In fact, several explosion models have been reported recently in
simulations that include multi-D effects that increase the neutrino
heating \citep{bura06,mare09,brue09}.  Based on the long-term
two-dimensional (2D) simulations with one of the best available
neutrino transport approximations, \citet{bura06} firstly report an
explosion for the non-rotating low-mass (11.2 $M_{\odot}$) progenitor
of \citet{woos02}. Due to the compactness of the iron core ($\sim 1.26
M_{\odot}$) with its steep outer density gradient, the explosion is
initiated at $\sim 300$ ms after core bounce. This is much earlier
than in \citet{mare09}, who observe the delayed onset of the explosion
$\sim 600$ ms for a $15 M_{\odot}$ progenitor of \citet{woos95} with a
moderately rapid rotation imposed.  Although the explosion mechanism
by neutrino-heating is very plausible, there are other possible
mechanisms, in which the magnetohydrodynamic (MHD) mechanism is
included (see references in \cite{kotarev,ober06,burr07b,taki09}).
Another suggested mechanism relies on acoustic energy deposition via
oscillating protoneutron stars (PNSs), which has been discovered by a
series of 2D multi-energy flux-limited-diffusion transport simulations
\citep{burr06,burr07a}. Although the additional energy input from
sound appears to be robust enough to explode even the most massive
progenitors \citep{burr07a}, it remains a matter of vivid debate and
has yet to be confirmed by other groups. Also exotic physics in the
core of the PNS may have a potential to trigger explosions (e.g.,
\cite{sagert}).

In this {\it Letter}, we present axisymmetric explosion models for a
13 $M_{\odot}$ progenitor model of \citet{nomo88} in support of the
theory that neutrino-heating aided by multi-D effects is able to cause
supernova explosions. We choose the progenitor with a smaller iron
core ($\sim 1.20 M_{\odot}$), anticipating an explosion since the
progenitor mass lies between 11.2 $M_{\odot}$ \citep{bura06} and $15
M_{\odot}$ \citep{mare09}.  We perform 2D core-collapse simulation
with spectral neutrino transport by the isotropic diffusion source
approximation (IDSA) scheme currently developed by \citet{lieb09}. By
comparing four exploding models with and without rapid rotation to one
non-exploding 1D model, we point out that models that produce a
north-south symmetric bipolar explosion can lead to larger explosion
energies than for the corresponding unipolar explosions. Our results
show that the explosion geometry is more likely to be bipolar in
models that include rotation.

\section{Numerical Methods and Models}

Our 2D simulations are performed using a newly developed code which
implements spectral neutrino transport using the IDSA scheme
\citep{lieb09} in a ZEUS-2D code \citep{ston92}.  Following the spirit
of the so-called ray-by-ray approach, the IDSA scheme further splits
the neutrino distribution into two components, both of which are
solved using separate numerical techniques. Although it does not yet
include heavy lepton neutrinos such as $\nu_{\mu}, \nu_{\tau}$
($\bar{\nu}_{\mu}, \bar{\nu}_{\tau}$) and the inelastic neutrino
scattering with electron, the innovative approach taken in the scheme
saves a significant amount of computational time compared to the
canonical Boltzmann solvers (see \cite{lieb09} for more details).
Expecting a bigger chance to produce explosions \citep{mare09}, we
employ the soft equation of state (EOS) by \citet{latt91} with a
compressibility modulus of $K=180$ MeV.  The self gravity is
implemented by solving the Poisson equation by the Modified Incomplete
Cholesky Conjugate Gradient (MICCG) method \citep{kota03b}, but
without relativistic corrections.

The simulations are performed on a grid of 300 logarithmically spaced
radial zones up to $5000$ km. To test the sensitivity with respect to
angular resolution, the grid is varied to consist of 64 or 128
equidistant angular zones covering $0 \leq \theta \leq \pi$.  For the
neutrino transport, we use 20 logarithmically spaced energy bins
reaching from 3 to 300 MeV.

All supernova calculations in this work are based on the $13 M_\odot$
model by \citet{nomo88}. The computed models are listed in the first
column of Table 1, in which one calculation (model M13-1D) is
conducted in spherical symmetry.  Other models are 2D simulations with
or without rotation (indicated by rot) with different numerical
resolution in the lateral direction (64 or 128, denoted by "hr" (high
resolution) in Table 1). For the rotating models, we impose rotation
on the progenitor core with initially a constant angular frequency of
$\Omega_0 = 2~{\rm rad}$/s inside the iron core with a dipolar cut off
($\propto r^{-2}$) outside, which corresponds to $\beta \sim 0.18 \%$
with $\beta$ being the ratio of the rotational to the gravitational
energy.  This rotation rate is fairly large and may lead to a spiral
mode of the SASI \citep{yama08}. In addition, this strong rotation may
induce a strong magnetic field due to winding and the
magneto-rotational instability and produce a jet-like outflow
\citep{kotarev}. Although these effects could modify the dynamics of
the postbounce phase, the approximate treatment in this study
(axisymmetry without magnetic fields) does not allow us to investigate
them in this article.

\begin{figure}[tbp]
\includegraphics[width=.4\textwidth]{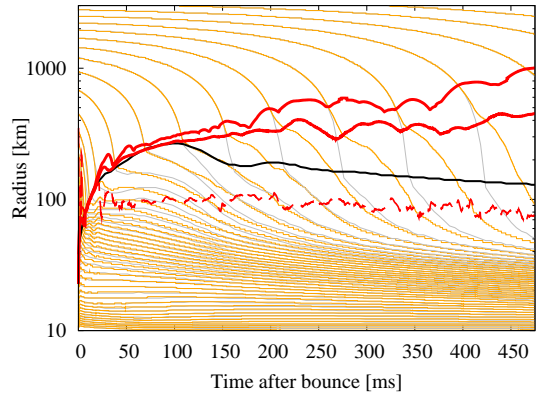}
\caption{Time evolution of Models M13-1D and M13-2D, visualized by
  mass shell trajectories in thin gray and orange lines, respectively.
  Thick lines in red (for model M13-2D) and black (model M13-1D) show
  the position of shock waves, noting for 2D that the maximum (top)
  and average (bottom) shock position are shown.  The red dashed line
  represents the position of the gain radius, which is similar to the
  1D case (not shown).}
\label{fig:1}
\end{figure}

\begin{figure*}[tbp]
\begin{center}
\includegraphics[width=.3\textwidth]{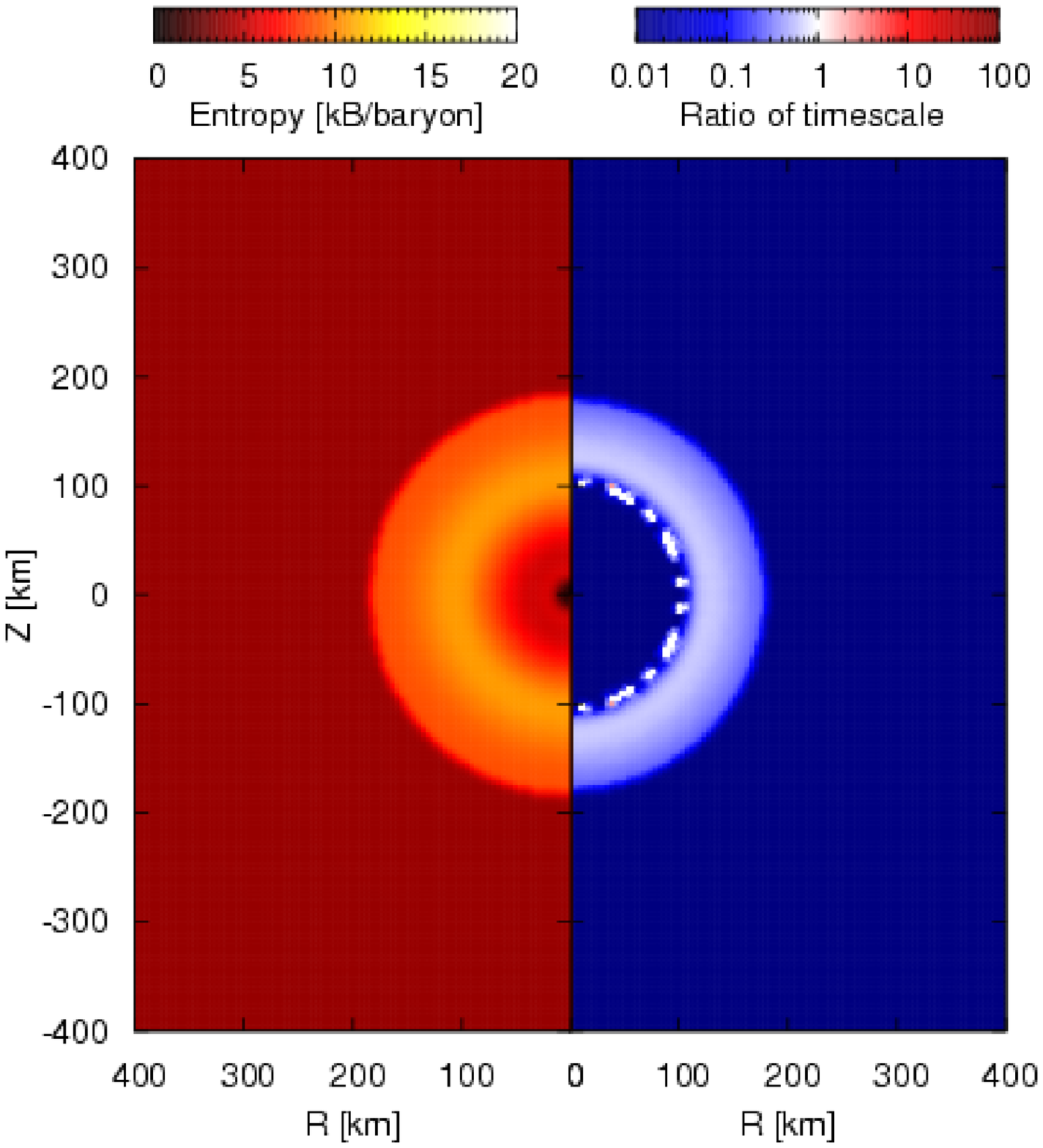}
\includegraphics[width=.3\textwidth]{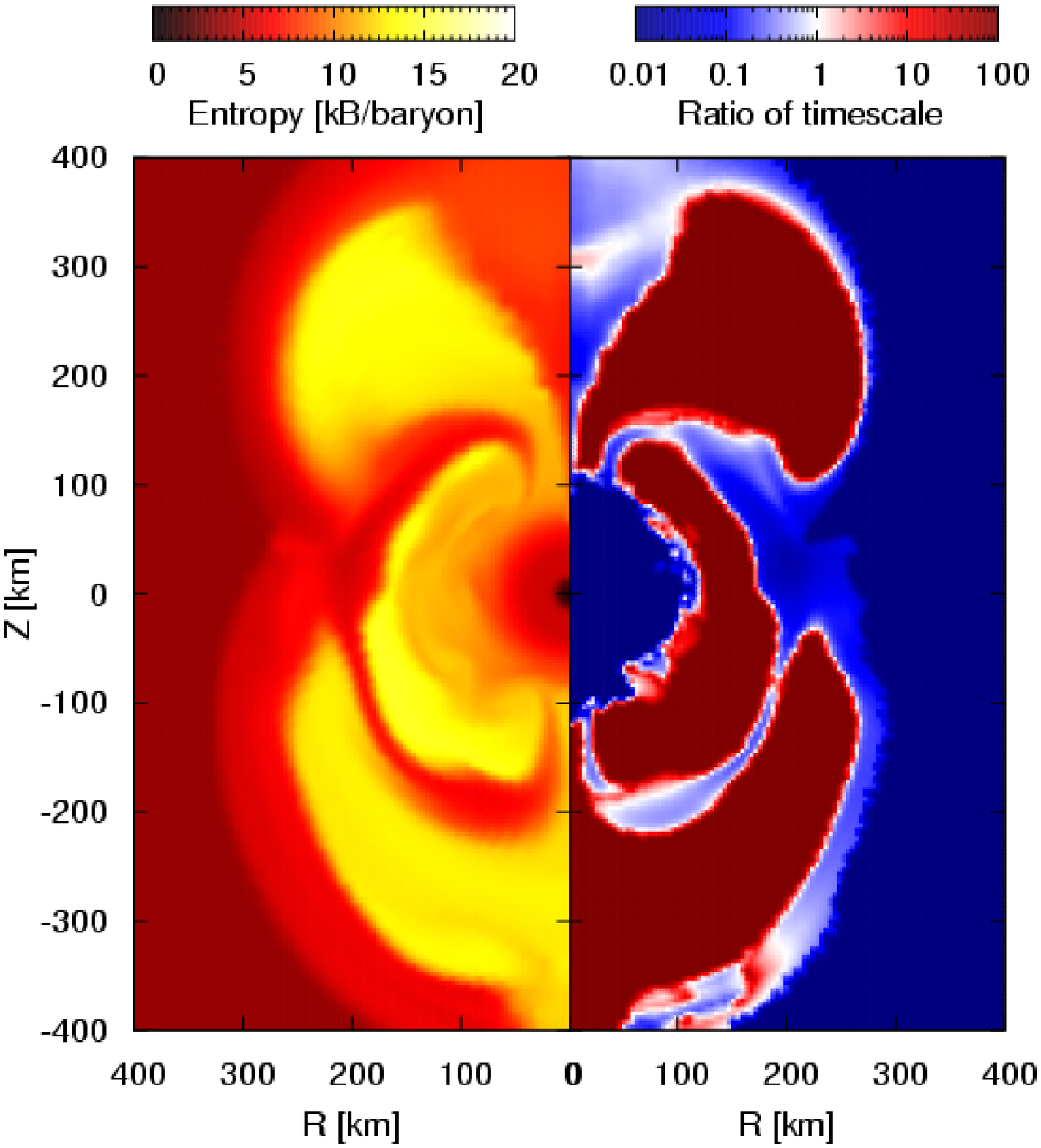}
\end{center}
\caption{Snapshot of the distribution of entropy (left half) and the
  ratio of the advection to the heating timescale (right half) for
  models of M13-1D (left) and M13-2D (right) at 200 ms after bounce.}
  \label{fig:2}
\end{figure*}

\section{Results}

Figure \ref{fig:1} depicts the difference between the time evolutions
of model M13-1D (thin gray lines) and model M13-2D (thin orange
lines), visualized by mass shell trajectories.  Until $\sim 100$ ms
after bounce, the shock position of the 2D model (thick red line) is
similar to the 1D model (thick black line). Later on, however, the
shock for model M13-2D does not recede as for M13-1D, but gradually
expands and reaches 1000 km at about 470 ms after bounce.  Comparing
the position of the gain radius (red dashed line) to the shock
position of M13-1D (thick black line) and M13-2D (thick red line), one
can see that the advection time of the accreting material in the gain
region can be longer in 2D than 1D.  This longer exposure of cool
matter in the heating region to the irradiation of hot outstreaming
neutrinos from the PNS is essential for the increased efficiency of
the neutrino heating in multi-D models.

A more detailed analysis of the timescale is shown in Figure
\ref{fig:2}. The right-half shows
$\tau_\mathrm{adv}/\tau_\mathrm{heat}$, which is the ratio of the
advection to the neutrino heating timescale.  For the 2D model (right
panel), it can be shown that the condition of
$\tau_\mathrm{adv}/\tau_\mathrm{heat}\gtrsim1$ is satisfied behind the
aspherical shock, which is deformed predominantly by the SASI, while
the ratio is shown to be smaller than unity in the whole region behind
the spherical standing accretion shock (left panel:1D). Note that
$\tau_\mathrm{heat}$ is estimated locally by $e_\mathrm{bind}/Q_\nu$,
where $e_\mathrm{bind}$ is the local specific binding energy (the sum
of internal plus kinetic plus gravitational energies) and $Q_\nu$ is
the specific heating rate by neutrinos, and that $\tau_\mathrm{adv}$
is given by $[r -r_\mathrm{gain}(\theta)]/|v_r(r,\theta)|$, where
$r_\mathrm{gain}$ is the gain radius and $v_r$ is the radial
velocity. Comparing the left-half of each panel, the entropy for the
2D model is shown to be larger than for the 1D model. This is also the
evidence that the neutrino heating works more efficiently in multi-D.

We now move on to discuss models with rotation. Both, for model
M13-rot and its high resolution counterpart, model M13-rot-hr, we
obtain neutrino-driven explosions (see, $t_{1000}$ and
$E_\mathrm{dia}$ in Table 1). The rapid rotation chosen for this study
mainly affects the explosion dynamics in the postbounce phase, which
we will discuss in the following.

\begin{figure*}[htbp]
\begin{center}
\includegraphics[width=.3\textwidth]{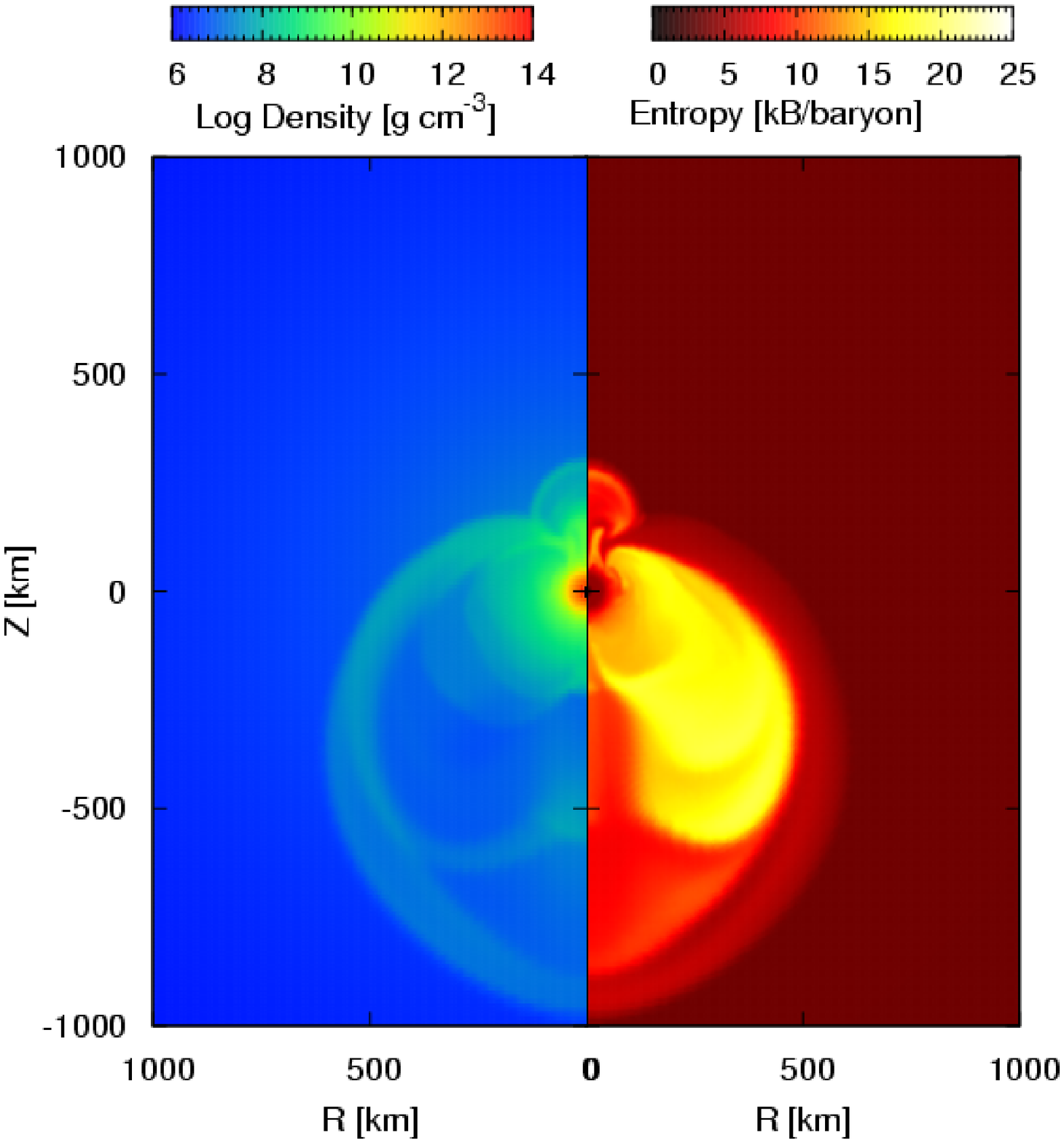}
\includegraphics[width=.3\textwidth]{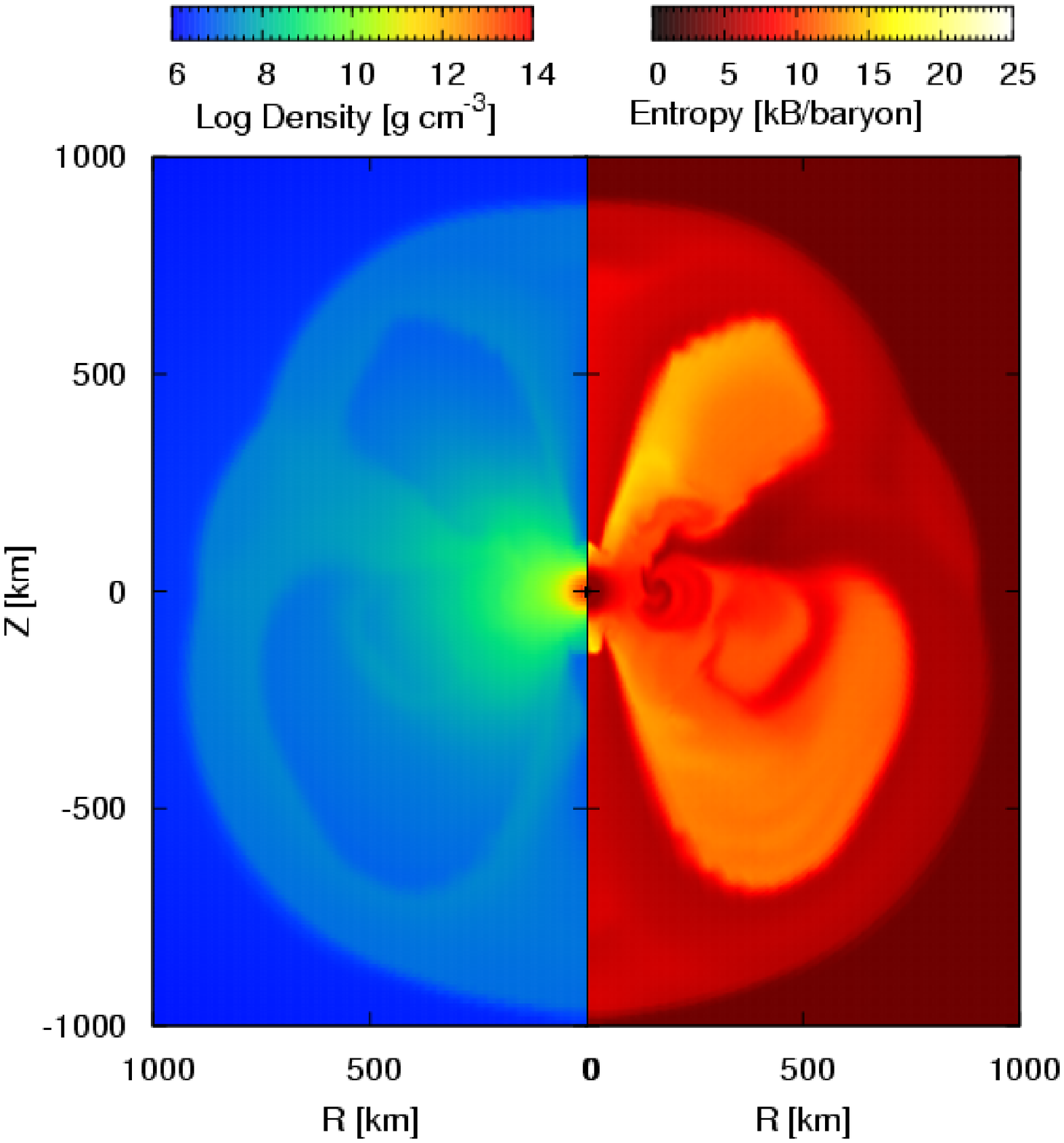}
\end{center}
\caption{Snapshots of the density (left half) and the entropy (right
  half) for models M13-2D (left) and M13-rot (right) at the epoch when
  the shock reaches 1000 km, corresponding to $\sim$ 470 ms after
  bounce in both cases.}
\label{fig:3}
\end{figure*}

For the rotating model, the dominant mode of the shock deformation
after bounce is almost always the $\ell =2$ mode although the $\ell =
1$ mode can be as large as the $\ell = 2$ mode when the SASI enters
the non-linear regime ($\gtrsim 200$ ms after bounce). In contrast to
this rotation-induced $\ell = 2$ deformation, the $\ell =1$ mode tends
to be larger than the $\ell = 2$ mode for the 2D models without
rotation in the saturation phase.  As shown in Figure \ref{fig:3},
this leads to different features in the shock geometry, namely the
preponderance of the unipolar explosion for the 2D models without
rotation (left), and the bipolar (north-south symmetric) explosion
with rotation (right).

\begin{figure}[h]
\centering
\includegraphics[width=.4\textwidth]{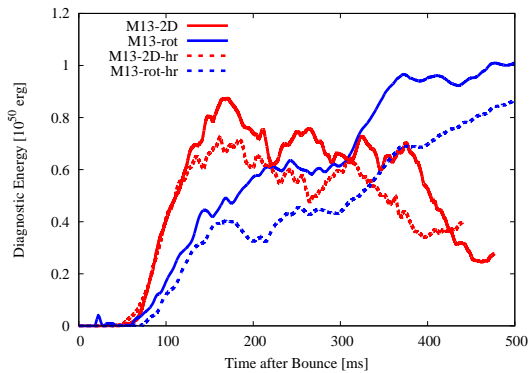}
\caption{Time evolution of the diagnostic energy versus postbounce
  time for 2D models with and without rotation. }
\label{fig:4}
\end{figure}

Since it is impossible to calculate precise explosion energies at this
early stage, we define a {\em diagnostic} energy that refers to the
integral of the energy over all zones that have a positive sum of the
specific internal, kinetic and gravitational energy.  Figure
\ref{fig:4} shows the comparison of the diagnostic energies for the 2D
models with and without rotation.  Although the diagnostic energies
depend on the numerical resolutions quantitatively, they show a
continuous increase for the rotating models. The diagnostic energies
for the models without rotation, on the other hand, peak around $180$
ms when the neutrino-driven explosion sets in (see also Figure 1), and
show a decrease later on. With values of order $10^{49}$ erg it is not
yet clear whether these models will also eventually lead to an
explosion.

The reason for the greater explosion energy for models with rotation
is due to the bigger mass of the exploding material.  This is because
the north-south symmetric ($\ell =2$) explosion can expel more
material than for the unipolar explosion.  In fact, the mass enclosed
inside the gain radius is shown to be larger for the rotating models
(e.g., Table 1). The explosion energies when we terminated the
simulation are less than $\lesssim 10^{50}$ erg for all the models.
For the rotating models, we are tempted to speculate that they could
become as high as $\sim 10^{51}$ erg within the next 500 ms by a
linear extrapolation. However, in order to unquestionably identify the
robust feature of an explosion in the models, a longer-term simulation
with improved input physics would be needed.

Our numerical results are qualitatively consistent with the results of
\citet{mare09} in the sense that in a relatively early postbounce
phase the model with rotation shows a more clear trend of explosion
than for the non-rotating models.

\section{Summary and Discussion}
By performing 2D core-collapse simulations of a 13 $M_{\odot}$ star
with spectral neutrino transport via the isotropic diffusion source
approximation, we have found a strong dependence of the expansion of
the shock radius and the likelihood for an explosion on the initial
rotation rate. In all cases the shock is driven outward by the
neutrino heating mechanism aided by multi-D effects such as the SASI
and convection.  We have shown the preponderance of an bipolar
explosion for 2D models with rotation. We have pointed out that the
explosion energy can become larger for models with bipolar explosions.

The conclusion with respect to the effects of rotation obtained in
this study differs from that of \citet{mare09}, who suggested that the
rotation has a negative impact on the explosion. They obtained the
expansion of the shock wave only for the rotating model (M15LS-rot),
while the nonrotating model did not show an expansion due to the short
simulation time (see Fig. 6 in their paper). Therefore they could not
compare the expanding shock evolution in both the rotating and the
nonrotating cases so that their discussion is limited to the shock
oscillation phase.

Here it should be noted that the simulations in this paper are only a
very first step towards more realistic supernova models (e.g.,
\cite{mare09,burr07a,brue09}).  The approximations adopted in this
paper should be improved, for example the omission of heavy lepton
neutrinos, the inelastic neutrino scattering, and the ray-by-ray
approach.  The former two, may act to suppress the explosion.  However
we think that qualitative effects induced by rotation will not be
affected so much because they are produced mainly by the hydrodynamic
interplay of the SASI and the rotation. The ray-by-ray approach may
lead to the overestimation of the directional dependence of the
neutrino anisotropies (see discussions in \cite{mare09}). On the other
hand, the lateral neutrino emission and the enhanced heating near the
polar regions, such as from the oblately deformed protoneutron star
due to rapid rotation (e.g., \cite{kota03b}), could be
underestimated. Apparently the full-angle transport will give us the
correct answer (e.g., \cite{ott08}). In addition, due to the
coordinate symmetry axis, the SASI develops preferentially along the
axis, thus it could provide a more favorable condition for the
explosion. As several exploratory simulations have been done recently
\citep{iwak08,sche08,iwak09}, 3D supernova models are indeed
necessary.

Bearing these caveats in mind, the role of rotation acting on the
neutrino-driven explosions, is qualitatively new. Yet there remain a
number of issues to be studied. We have to clarify the progenitor
dependence and also investigate the effects of rotation more
systematically by changing its strength in a parametric manner
(possibly with magnetic fields). It will be interesting to study the
neutrino and gravitational-wave signals. This paper is a prelude for
the forthcoming work that will clarify these issues one by one.

\bigskip
Y.S would like to thank to E. M\"uller and H.-Th. Janka for their kind
hospitality during his stay in MPA.  K.K is grateful to S. Yamada for
continuing encouragements.  Numerical computations were in part
carried on XT4 at CfCA of the National Astronomical Observatory of
Japan. S.C.W and M.L are supported by the Swiss National Science
Foundation under grant No. PP00P2-124879 and 200020-122287.  This
study was supported in part by the Japan Society for Promotion of
Science (JSPS) Research Fellowships (YS) and the Grants-in-Aid for the
Scientific Research from the Ministry of Education, Science and
Culture of Japan (Nos. 19540309 and 20740150).


\begin{table*}[htbp]
  \begin{center}
  \caption{Model summary}
  \begin{tabular}{c|cccccc}
    \hline
    Models & Dimension & $\Omega_\mathrm{0}$ &  $N_\theta$ & $t_\mathrm{1000}$ 
    & $E_\mathrm{dia}$  & $M_\mathrm{gain}$\\ 
    &  & [rad/s] & & [ms] & [$10^{50}$ erg] & [$M_\odot$]\\
    \hline\hline
    M13-1D     & 1D & -- & 1 & -- & -- & -- \\
    M13-2D & 2D & -- & 64 & 470 & 0.26 & 0.017\\
    M13-rot   &2D \& rotation & 2 & 64 & 480 & 0.95 & 0.067\\
    M13-2D-hr &2D & -- & 128 & 420 & 0.40 & 0.018\\
    M13-rot-hr  &2D \& rotation & 2 & 128 & 520 & 0.78 & 0.060\\
    \hline
    \multicolumn{1}{@{}l@{}}{\hbox to 0pt{\parbox{120mm}{\footnotesize
          \footnotemark[$$] 
          $\Omega_0$ is the precollapse angular velocity.
          $N_\theta$ represents the lateral grid number covering $0 \leq
          \theta \leq \pi$.  ``hr (high resolution)'' indicates the runs for
          $N_{\theta} = 128$. $t_\mathrm{1000}$ represents the time
          (measured after bounce) when the average shock radius becomes 1000
          km. $E_\mathrm{dia}$ is the diagnostic energy defined as the total
          energy (internal plus kinetic plus gravitational), integrated over
          all matter where the sum of the corresponding specific energies is
          positive.  $M_\mathrm{gain}$ is the mass inside the gain
          layer. The latter quantities are given at 450 ms postbounce.
        }\hss}}
  \end{tabular}
  \label{tab1}
  \end{center}
\end{table*}

\end{document}